\documentclass[journal=jacsat,manuscript=article]{achemso}

\usepackage{chemformula} % Formula subscripts using \ch{}
\usepackage[T1]{fontenc} % Use modern font encodings
\usepackage{amssymb,amsmath,mathtools}

\author{Arnab Barman Ray}
\affiliation{The Institute of Optics, University of Rochester}
\author{Kevin Liang}
\affiliation{Adelphi University}
\author{Nick Vamivakas}
\email{nick.vamivakas@rochester.edu}
\phone{+1(585) 275-2089}
\affiliation{The Institute of Optics, University of Rochester}

\title[An \textsf{achemso} demo]
  {Valley engineering electron-hole liquids in TMDC monolayers}

\abbreviations{IR,NMR,UV}

\begin{document}

\begin{tocentry}

Some journals require a graphical entry for the Table of Contents.
This should be laid out ``print ready'' so that the sizing of the
text is correct.

Inside the \texttt{tocentry} environment, the font used is Helvetica
8\,pt, as required by \emph{Journal of the American Chemical
Society}.

The surrounding frame is 9\,cm by 3.5\,cm, which is the maximum
permitted for  \emph{Journal of the American Chemical Society}
graphical table of content entries. The box will not resize if the
content is too big: instead it will overflow the edge of the box.

This box and the associated title will always be printed on a
separate page at the end of the document.

\end{tocentry}

\begin{abstract}
	
	Electron-hole liquids(EHLs), a correlated state of matter and a thermodynamic liquid, have recently been found to exist at room temperature in suspended monolayers of $\text{MoS}_2$.
	Appreciably higher rates of radiative recombination inside the liquid as compared to free excitons hold promise for optoelectronic applications such as broadband lasing. In this paper, we show that leveraging the valley physics in $\text{MoS}_2$ may be a route towards achieving tunability of specific characteristics of an EHL, such as emission wavelength, linewidth, and most importantly, the liquid density. The conditions under which EHLs form, in bulk semiconductors as well as TMDC monolayers are quite stringent, requiring high crystal purity and cryogenic temperatures in bulk semiconductors, and suspension in monolayers. Using a simple yet powerful model for describing free excitons and show that a phase transition into the EHL state may be feasible in substrate-supported monolayer samples. More repeatable experimental realizations of EHLs may be essential to answer questions regarding the nature of electron-hole correlations and how they may be used to generate non-trivial states of light. 
  
\end{abstract}

\section{Introduction}

The characteristic broad luminescence spectra associated with an electron hole liquid(EHL) was first detected by J.R. Haynes at the Bell Telephone Laboratories in 1966, in a sample of silicon immersed in liquid helium, when the pump intensity exceeded a certain threshold\cite{1}. Subsequent experiments in the late 60s and early 70s provided further evidence for EHLs. It was L.V. Keldysh who first provided a theoretical foundation for a phase transition, a gas-liquid condensation in semiconductors at sufficiently low temperatures, where the interaction between electron and holes at large densities ($~10^{17} \text{cm}^{-3}$) would cause them to condense from an insulating gas of free excitons and possibly multi-excitonic complexes into macroscopic metallic droplets of delocalized electrons and holes\cite{2,3}. In Ge, experiments showing abrupt increases in photo-conductivity\cite{4} with laser pump intensity and collection of charge pulses in p-n junctions at high incident intensities\cite{5} further established the distinctness of a metallic EHL from candidates like a dielectric $H_2$-like excitonic liquid or a BEC. Infrared scattering experiments showed the state existing as a fog of smaller droplets while large $\mu$m-sized droplets were found to form under certain conditions\cite{6}. Theory and experiments agreed with each other, and while open questions remained, the phenomenon faded into obscurity until recently when EHLs were realized at room temperatures in atomically thin semiconductors. Room temperature realization of these macroscopic quantum states has revived interest and relevance\cite{7,8,9,10,11}.

Using a suspended CVD-grown monolayer of $\text{MoS}_2$, kept at room temperatures, a sharp transition in the photoluminescence spectrum was recently observed\cite{8}. The A exciton line disappears into a broad, red-shifted peak at high power densities, accompanied by a corresponding reduction in the emission spot size. This transition was shown to be reversible, ruling out defect-generation as a cause for the observation. Theoretical calculations for EHL formation in 2-dimensional TMDCs as well as thermodynamic analyses that use the latest standards for quantifying carrier interactions in TMDCs, with the use of the Keldysh potential, do show the possibility of such a transition for temperatures above $500$ K\cite{9,11}. Another evidence for the formation of an EHL has surfaced in graphene-encapsulated $\text{MoTe}_2$, in the form of abrupt changes in photocurrent across the monolayer at high laser intensities\cite{10}.

In experiments, the use of a suspended monolayer is crucial for two reasons. First, due to the reduced dielectric screening compared to substrate-supported samples, Coulomb interactions are stronger. Secondly, the lack of a constraining substrate allows the monolayer to undergo isotropic expansion, affecting a transition into an indirect band-gap semiconductor, imparting larger lifetimes to charge carriers and allowing them to reach a quasi-equilibrium state. This paper demonstrates the possibility of leveraging the unique spin-valley physics of TMDC monolayers to tune the EHL emission properties, the binding energy, emission linewidth, and liquid density. Using a simple model that captures the strength of electromagnetic interactions in monolayers, we show that even with dielectric screening from a substrate, as long as the charge carriers stay in the system long enough, with lifetimes exceeding quasi-thermalization times\cite{2}, a phase transition into an EHL is theoretically feasible.

\section{Valley Engineering an EHL}

\subsection{Binding Energy}

\begin{figure}[h]
\centering
\includegraphics[height=0.6\textwidth,width=0.7\textwidth]{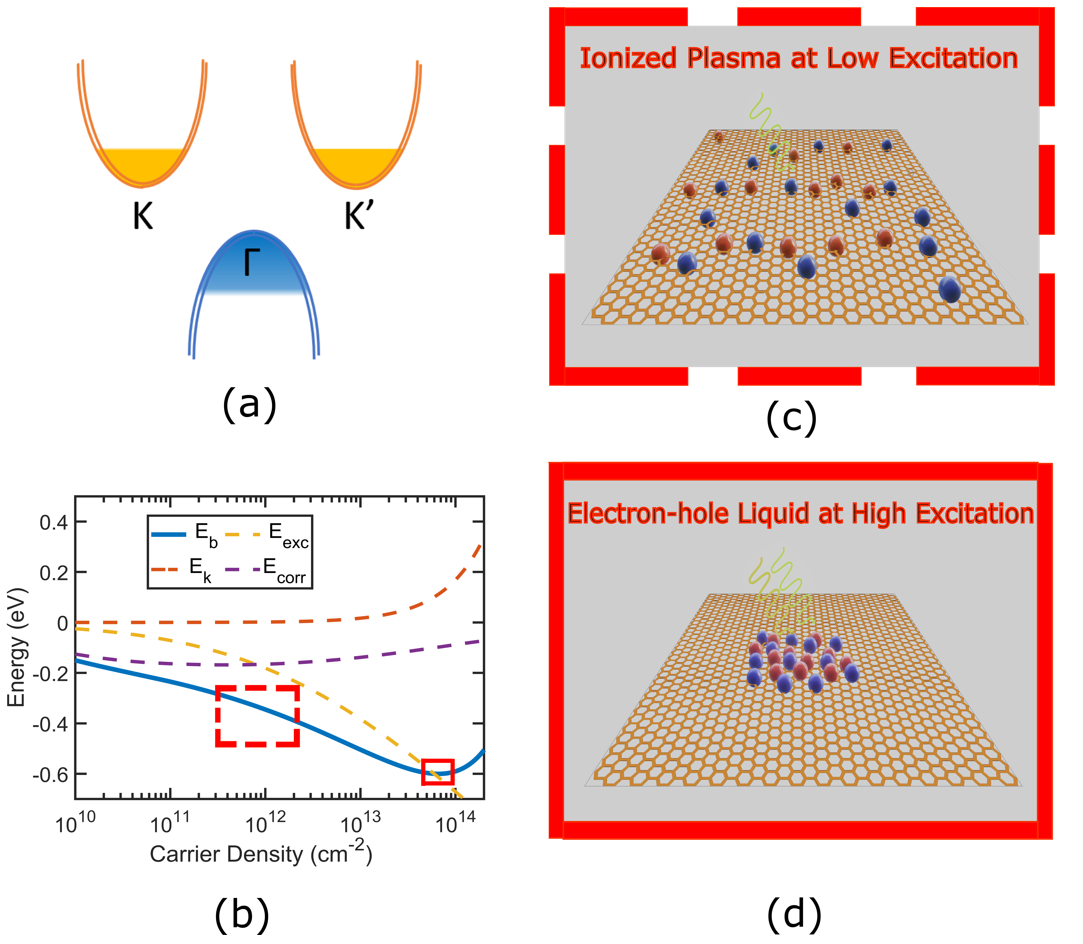}
\caption{(a) Band energy diagram of monolayer $\text{MoS}_2$, filled portions indicate the amount of charge carriers in a particular valley, (b) Binding energy with carrier density -  dashed and solid squares designate regions of ionized plasma vs EHL, as illustrated in (c) and (d) respectively.}
\label{fig1}
\end{figure}

The binding energy of an electron-hole pair in an EHL is found from the minima of the binding energy for a correlated plasma gas, consisting of electrons and holes, as the carrier density is varied. The binding energy of such a system consists of three parts, $E_b = E_k+E_{exc}+E_{corr}$, as a sum of the kinetic, exchange and correlation energies. In keeping with the current understanding that an EHL arises out of an indirect bandgap transition for a monolayer of $\text{MoS}_2$, the system consists of 2 nearly spin-degenerate electron bands at the $K$ and $K^\prime$ points of the first Brillouin zone, and a spin-degenerate hole band at $\Gamma$, as shown in Fig.~\ref{fig1}(a). It has been shown that the majority of holes reside at $\Gamma$ at high carrier densities\cite{12}.
In the following, we calculate the binding energy in a manner similar to Rustagi et.al\cite{9}. 

The first contribution to the energy per electron-hole (e-h) pair is the ground state, non-interacting kinetic energies of the electrons and holes. For the case of our model, we have
\begin{equation}
  E_k = \frac{2\pi\hbar^2}{N_{eh}}(\int_{0}^{k_{eK}}dk\,\frac{k^3}{m_e} +\int_{0}^{k_{eK^{\prime}}}dk\,\frac{k^3}{m_e}+\int_{0}^{k_{h\Gamma}}dk\,\frac{k^3}{m_h})\, ,  
\end{equation}
where $k_{eK}$,$k_{eK^{\prime}}$ and $k_{h\Gamma}$ are the Fermi wavevectors of the three valleys. For equal valley contributions, $k_{eK} = k_{eK^{\prime}} = \sqrt{\pi N_{eh}}$ and $k_{h\Gamma} =  \sqrt{2\pi N_{eh}}$, where $N_{eh}$ is the density of e-h pairs.

The exchange energy, that comes out of the requirement for anti-symmetry in the many body wavefunction is simply the expectation value of the interaction potential in the ground state of the system and is given as:
\begin{multline}
E_{exc} = -\frac{1}{(2\pi)^2N_{eh}}(\int\int_{F_{K}}d\mathbf{k}\mathbf{k^\prime}v(|\mathbf{k}-\mathbf{k^{\prime}}|)+ \\ \int\int_{F_{K^{\prime}}}d\mathbf{k}\mathbf{k^\prime}v(|\mathbf{k}-\mathbf{k^{\prime}}|)+ \int\int_{F_{\Gamma}}d\mathbf{k}\mathbf{k^\prime}v(|\mathbf{k}-\mathbf{k^{\prime}}|))\, ,
\end{multline}
where $v(|\mathbf{q}|) = \frac{1}{L^2}\frac{e^2}{2\epsilon_0(\epsilon_1+\epsilon_2)|\mathbf{q}|(1+r|\mathbf{q}|)}$, with screening length\cite{13} $r = 4\pi r_0/(\epsilon_1+\epsilon_2)$ where $r_0$ is the polarizability in TMDC monolayers\cite{14}. The integrals are within the respective Fermi surfaces of the three valleys. Analytically, each of the three terms in Eq.(2), can be represented as $E_{k_{F}} = E_1\, +\, E_2$, where $E_1 = -\frac{4 k_F^2}{3\pi N_{eh}}$, and $E_2 = -\frac{2\pi k_F^2}{N_{eh}} \int_{-\infty}^{\infty}\frac{r}{\pi^2x^2}\, G^{1,3}_{3,1}[4(\frac{r}{x})^2|\begin{smallmatrix} 0, 0, \frac{1}{2} \\ \frac{1}{2} \end{smallmatrix}]\, (\frac{J_1(k_F\,x)}{x})^2\,x\,dx$, for a Fermi wave-vector $k_F$. The G-function here is the Meijer G-function. 

The correlation energy is the perturbative correction that arises from turning on the interaction between the fermions. It's related to the random-phase approximation (RPA) dielectric function, $\epsilon_{RPA} = 1 - a_0(\mathbf{k},\omega)-i\sigma_0(\mathbf{k},\omega)$ through\cite{15}:
\begin{equation}
E_{corr} = -\frac{1}{4\pi N_{eh}}\int\frac{d\mathbf{k}}{(2\pi)^2}\int\hbar\, d\omega\{sgn(\sigma)\,\frac{\sigma_0}{\sigma}\,\tan^{-1}(\frac{|\sigma|}{1-a})-\sigma_0\}\,.
\end{equation}

For a single spin-degenerate band, the polarizabilities, $\sigma_l$ and $a_l$ can be calculated as: $ a_l +i\sigma_l = -2v(|\mathbf{k}|)\sum_q\frac{n_{k+q}-n_p}{\hbar\omega+E(\mathbf{q})-E(\mathbf{q}+\mathbf{k})}$ - explicit values of which for isotropic 2D bands have been previously calculated\cite{16}. The total polarizabilities $\sigma_0$ and $a_0$ are simply the sum of the respective terms for the $3$ valleys. The exchange corrected values of the polarizabilities, which arise out of the reduced correlation between electrons of the same spin due to Pauli exclusion are obtained as\cite{17}:
$\sigma = (1-h_{e1}(k))\sigma_{0}^{e1}+(1-h_{e2}(k))\sigma_{0}^{e2}+(1-h_{h}(k))\sigma_{0}^{h}$, where $h(k) = v(\sqrt{k^2+f^2})/2v(k)$ is the correction term for a valley with fermi wavevector $f$. The real part of the corrected polarizability $a$ is given by, $a = \frac{\sigma}{\sigma_0}a_0$.

We chose values of the conduction band effective mass from ab-initio studies, $m_e = 0.55$\cite{RN115,RN116}, and the effective mass in the $\Gamma$ Valley is taken to be $m_h = 2.0$, in order to obtain agreement with reported values of the linewidth, or the energy-edge width, as we show next. This value of the mass is consistent with a computational study of valley masses in strained and unstrained monolayers\cite{RN33}.

The binding energy is shown in Fig.~\ref{fig1}(b) as a function of carrier density. The minima at around $n_{ehl}= 6.5 \times 10^{13}\,\text{cm}^{-2}$ imparts the plasma with the properties of a thermodynamic liquid, with a well-defined density and volume. We note that the liquid density calculated here is higher than previous reports\cite{8}, as we believe that experimentally obtained values of EHL density are under-estimates, as EHL droplets exist in equilibrium with surrounding plasma inside the emission spot. The binding energy is estimated to be, $E_{b}^{ehl} = -0.61\, \text{eV}$. The energy-edge width, i.e the difference between the high and low energy tails of EHL emission would be the total of the Fermi energies in the CB and VB given by: $\Delta_{E} = E_{Kf}\,+\,E_{\Gamma f} + 2\,k_B\,T = 0.27\,\text{eV}$, also consistent with the values observed experimentally (we use $T = 460 K$). We note that these binding energies are orders of magnitude higher than those reported in bulk semiconductors at cryogenic temperatures owing to the strength of interactions in monolayers\cite{RN4}. Such high binding energies allow the EHL to be observed experimentally at room temperatures.

\subsection{Valley-induced EHL modulation}

The selection rules in monolayer $\text{MoS}_2$ allow the charge carriers to have a net valley polarization depending on the polarization of the excitation laser. Using elliptical or circular polarization, one can induce a valley polarization in the population of charge carriers. This allows the possibility of establishing a knob on the average kinetic energy of the total carrier population. While both the exchange and correlation energies are affected, the kinetic energy is the most sensitive to valley polarizing carriers. We note that the optical transitions are direct and still across the valleys at $K$ and $K^\prime$, allowing a valley polarization, even though most holes reside at $\Gamma$ after scattering at steady state. 

\begin{figure}[t]
	\includegraphics[height=0.6\textwidth,width=0.7\textwidth]{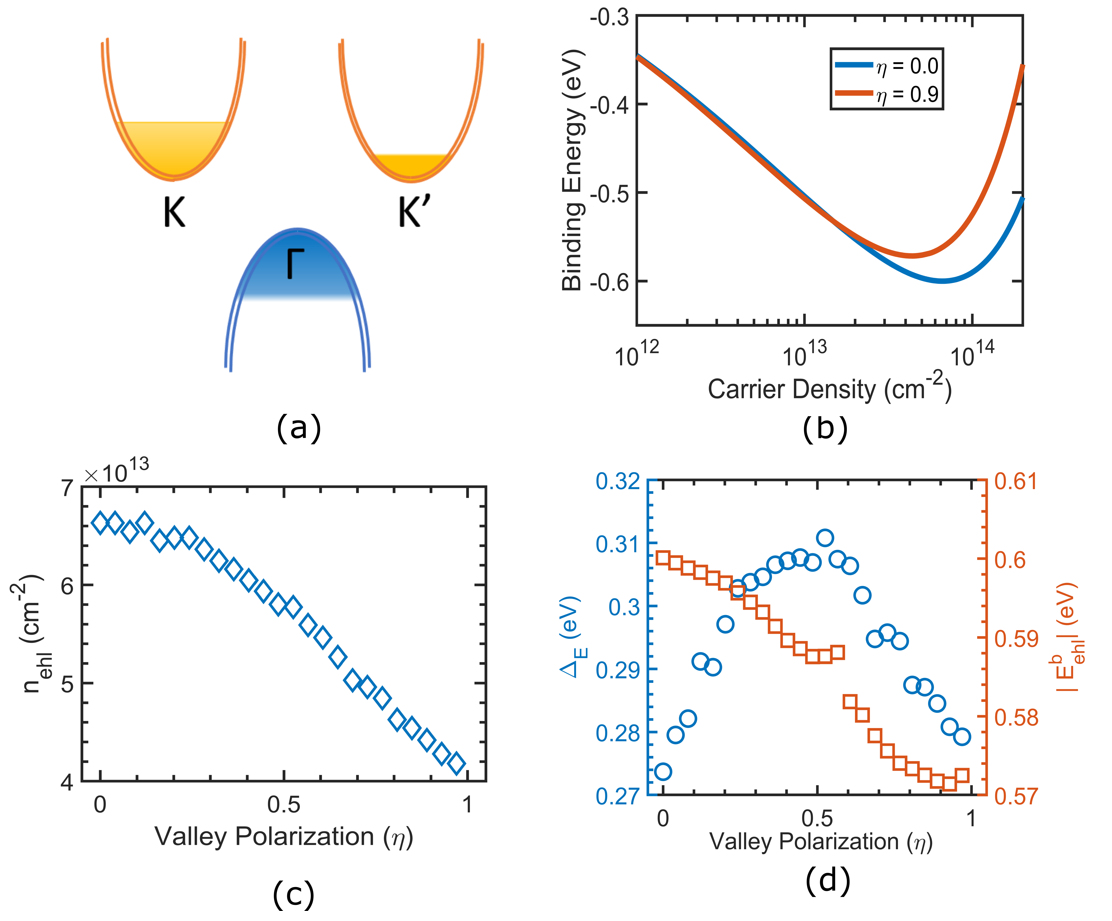}
	\caption{(a) Band diagram illustrating valley polarization with unequal number of charge carriers in each valley, (b) binding energy diagrams for $\eta = 0.0$ and $\eta = 0.9$, showing change with valley polarization, (c) EHL density with $\eta$ and (d) EHL binding energy and emission linewidth with $\eta$. }
	\label{fig2}
\end{figure}

In Fig.~\ref{fig2}(b), we see that while the EHL binding energy is mostly unaffected, the liquid density changes by an appreciable amount. The valley polarization is defined as $$\eta = \frac{N_{ehK}-N_{ehK^{\prime}}}{N_{ehK}+N_{ehK^{\prime}}}\, ,$$ where $N_{eh}=N_{ehK}+N_{ehK^{\prime}}$ and the Fermi wave-vectors for the 2 valleys are given as: $k_{eK}= \sqrt{2\pi\,\frac{1-\eta}{2}N_{eh}} $ and $k_{eK^{\prime}}= \sqrt{2\pi\,\frac{1+\eta}{2}N_{eh}} $. We note that while a value of $\eta=0.9$ is unrealistic in monolayers even at cryogenic temperatures, there have been studies indicating $100\,\%$  polarization in $\text{MoS}_2$\cite{18} with resonant excitation.

Fig.~\ref{fig2}(c) shows that $n_{ehl}$ is tunable over a wide range of $6.5\times 10^{13}\, \text{cm}^{-2}$ to $4 \times 10^{13}\,\text{cm}^{-2}$. As the valley polarization is increased, the average kinetic energy of the system increases, causing the liquid to be more destabilized, and decreasing its density. While the line-width of emission, now defined as $\Delta_{E} = \text{max}(E_{Kf},E_{K^{\prime}f})\,+\,E_{\Gamma f}$ should increase continuously as the Fermi energy in one of the valleys keeps increasing, we find that due to the lowering liquid density, the two effects nearly compensate each other and the line-width starts to decrease for $\eta\,>\,0.6$ as shown in Fig.~\ref{fig2}(d). Hence we can achieve around 10$\%$ increase in the emission linewidth. The change in binding energy is not appreciable within the studied range, however, due to band gap renormalization effects at different carrier densities, one may expect the EHL emission to shift, or even become bi-modal. This, however, is outside the scope of this work.

\section{Feasibility of an EHL in MgO-capped Samples}

Suspended monolayers are hard to fabricate owing to their intrinsic fragility. Furthermore, there is the added complication of laser heating. At higher laser powers, of the order of a few $\text{kW/cm}^2$, the monolayer experiences isotropic strain around $ \sim \,1.25\%$  to become an indirect bandgap semiconductor - which is essential for the charge carriers to remain in the system long enough to reach a quasi-equilibrium state. Recently, using thermally deposited oxide films, it has been demonstrated that high degrees of isotropic strain can be imparted to $\text{MoS}_2$ monolayers\cite{19}. We investigate the dielectric feasibility of EHL formation in such kinds of samples with compressive thin films of $\text{MgO}$ (IR dielectric constant: $2.6$), where the monolayer is supported on an hBN layer.

We start with a simple Wannier-Mott model of 2D excitons in TMDC monolayers that allow us to accurately capture the effect of dielectric screening and carrier-carrier interactions (see Supplementary Information for details). Figure ~\ref{fig3}(a) shows the ranges of calculated binding energies for different monolayer-substrate combinations and corresponding experimentally reported values\cite{RN1,RN2,RN3,RN4,RN5}. Since the experimental values of carrier densities are inaccessible, we use a constant and experimentally accessible carrier density range of $10^{11}-10^{12} \quad \text{cm}^{-2}$. We see that the model offers excellent qualitative agreement without the need for any fitting parameters. IR values of the dielectric constants are used in the calculation and are mentioned in the figure. Figure ~\ref{fig3}(c) demonstrates how the tunability of an EHL changes with increased environmental dielectric screening. We see that while the EHL density decreases with substrate screening, the degree of density modulation $\sim 40\%$ remains mostly constant.

\begin{figure}[h]
	\centering
	\includegraphics[height=0.6\textwidth,width=0.7\textwidth]{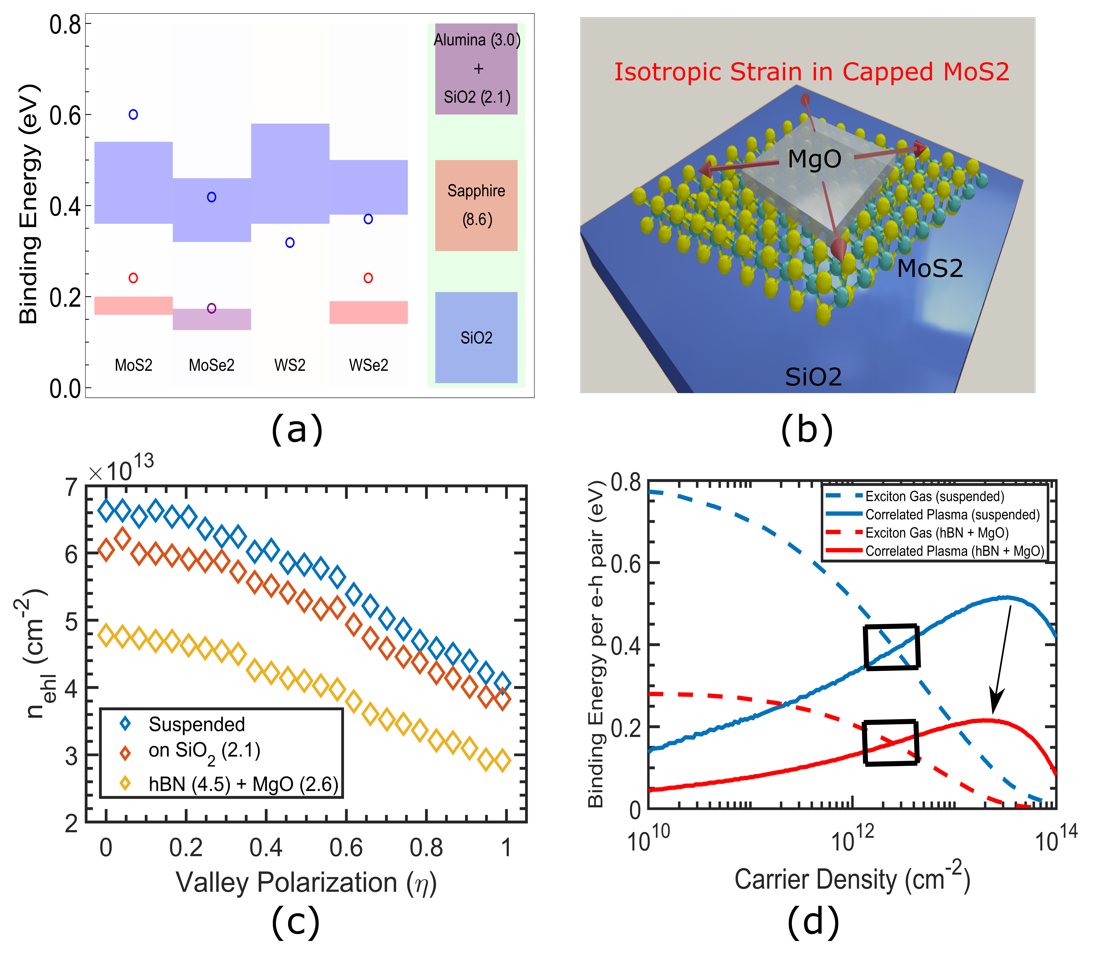}
	\caption{(a) A comparison of calculated(shaded regions) and reported binding energies(open circles) for excitons, (b) schematics of capped monolayer MoS2 samples, (c) EHL density tuning for different substrate super-strate configuration, (d) exciton gas vs EHL competition for suspended and MgO capped samples. }
	\label{fig3}
\end{figure}

A transition from the excitonic response at low carrier densities is required for an EHL to form.  While a complete thermodynamic analysis may be essential to ascertain whether a phase transition is feasible in the presence of environmental dielectric screening in substrate-supported samples, our results for the EHL and excitonic gas can be combined to show that dielectric screening is not an impediment to the realization of an EHL state in MgO-capped samples, as shown in Fig.~\ref{fig3}(d). The black arrow indicates the movement of the EHL state to lower densities. The black rectangles identify the regions of exciton-EHL crossover and illustrate no qualitative change in the nature of competition between these two states. We also note that the calculated binding energy of $\sim 0.2 \,\text{eV}$ for an EHL guarantees stability at room temperatures.

\section{Discussion}

As we show here, valley engineering opens up the possibility of tuning the characteristics of EHL emission. To the best of our knowledge, this is the first demonstration of its kind exploring the tunability of the liquid density with incident polarization. The tunable liquid density may find applications in generating and detecting THz waves. Using a highly successful free exciton model, we accurately capture the physics of dielectric screening and carrier interactions and apply it to assert the possibility of an EHL state forming in substrate-supported samples, even in the presence of dielectric screening. Hence we conclude that as long as a monolayer strains into an indirect bandgap, allowing charge carriers enough time to thermalize, an EHL state should be achievable.

High rates of radiative recombination inside a droplet ensure better PL quantum yield without the use of post-processing techniques\cite{RN16,RN23}. This holds promise for harnessing such collective phenomena for optoelectronic applications. More accessible samples allowing highly repeatable realizations of EHLs are essential to answer some pertinent questions. The nature of the quantum mechanical electron-hole correlations present in the liquid - which, through quantum statistical state injection\cite{20,RN28}, may be transferred into the emitted photons - can allow perhaps, the creation of non-trivial quantum states of light. Delocalized holes and electrons across a droplet may give rise to spatially coherent emission. There are present also interesting questions of in-liquid mobility and resistance. Our results provide fresh and fertile ground for further experimental exploration.

\begin{acknowledgement}

The author thanks Arunabh Mukherjee for discussions about the manuscript.

\end{acknowledgement}

\begin{suppinfo}
Extensive description of the Wannier Mott model used in this paper can be found in Supplementary Information.
\end{suppinfo}

\bibliography{achemso-demo}

\end{document}

We employ a simple variational method to obtain a Wannier-like Equation which describes A excitons in TMDC monolayers\cite{19}. While many sophisticated models of excitons have allowed tremendous insight  into the excitonic landscape of TMDCs\cite{20,21,22,23}, the results obtained here illustrate the robustness of our spinless, simplistic approach and how it successfully captures the physics of dielectric screening and carrier-carrier interactions, without the need for any fitting parameters.

\subsection{System Hamiltonian}

\begin{flushleft}
	We start with the Hamiltonian consisting of the kinetic energy term and the inter-carrier Coulomb interaction in the form of the Keldysh Potential\cite{13},
	
	\begin{align*}
	\hat{H} = &\sum_{\mathbf{k},\gamma}\epsilon_{\mathbf{k},\gamma}a^{\dagger}_{\gamma,\mathbf{k}}a_{\gamma,\mathbf{k}}\,+\\ \,\frac{1}{2}&\sum_{\mathbf{k},\mathbf{k^\prime},\mathbf{q},\gamma,\gamma^\prime}V_\textbf{q}\,a^{\dagger}_{\gamma,\mathbf{k}+\mathbf{q}}a^{\dagger}_{\gamma^\prime,\mathbf{k^\prime}-\mathbf{q}}a_{\gamma^\prime,\mathbf{k^\prime}}a_{\gamma,\mathbf{k}}
	\end{align*}
	
	where $V_{\mathbf{q}} = \frac{1}{L^2}\frac{e^2}{2\epsilon_0(\epsilon_1+\epsilon_2)|\mathbf{q}|(1+r|\mathbf{q}|)}$, with screening length $r = 4\pi r_0/(\epsilon_1+\epsilon_2)$ where $r_0$ is the polarizability in TMDC monolayers. Note that the potential is in MKS units. The kinetic energies are described by $\epsilon_{\mathbf{k},\gamma} = \frac{\hbar^2k^2}{2m_\gamma}$ for the electrons(holes). 
	The different bands in the first Brillouin zone(FBZ) of the monolayers are represented through their labels $\gamma \epsilon \{c,c^{\prime},v\}$, for the two nearly degenerate conduction bands and the topmost valence band. The model treats dark, spin-forbidden and bright excitons on an equal footing.
	
	In the absence of inter-valley correlations, which is a reasonable assumption at low carrier densities, only the bands at a single valley may be considered for the calculation.
	
	For a many body state ${\psi}_{MB}$, using a cluster expansion, the Hamiltonian can be used to give an expression for the total energy in terms of carrier distributions $f^{\gamma}_\mathbf{k} = \braket{a^{\dagger}_{\gamma,\mathbf{k}} a_{\gamma,\mathbf{k}}}$ and electron-hole, electron-electron and hole-hole correlations $c^{\mathbf{q},\mathbf{k},\mathbf{k^{\prime}}}_{eh}= 4\Delta \braket{a^{\dagger}_{c,\mathbf{k}}a^{\dagger}_{v,\mathbf{k^\prime}}a_{c,\mathbf{k^\prime}}a_{v,\mathbf{k}}}$, $c^{\mathbf{q},\mathbf{k},\mathbf{k^{\prime}}}_{ee} =4\Delta \braket{a^{\dagger}_{c,\mathbf{k}}a^{\dagger}_{c,\mathbf{k^\prime}}a_{c,\mathbf{k^\prime}}a_{c,\mathbf{k}}} $ and $c^{\mathbf{q},\mathbf{k},\mathbf{k^{\prime}}}_{hh} = \Delta \braket{a^{\dagger}_{v,\mathbf{k}}a^{\dagger}_{v,\mathbf{k^\prime}}a_{v,\mathbf{k^\prime}}a_{v,\mathbf{k}}}$:
	
	\begin{multline}
	E_{MB} = \sum_{\mathbf{k}}(\frac{\hbar\mathbf{k}^2}{m_e}f^{e}_{\mathbf{k}}+\frac{\hbar\mathbf{k}^2}{2m_h}f^{h}_{\mathbf{k}}) - \frac{1}{2}\sum_{\mathbf{k},\mathbf{k^{\prime}}}V_{\mathbf{k}-\mathbf{k}^{\prime}}(4f^{e}_{\mathbf{k}}f^{e}_{\mathbf{k}^{\prime}}\\ + f^{h}_{\mathbf{k}}f^{h}_{\mathbf{k}^{\prime}}) + \frac{1}{2}\sum_{\mathbf{k},\mathbf{k^{\prime}},\mathbf{q}}(V_{\mathbf{q}}(c^{\mathbf{q},\mathbf{k},\mathbf{k^{\prime}}}_{ee}+c^{\mathbf{q},\mathbf{k},\mathbf{k^{\prime}}}_{hh})-2V_{\mathbf{k}^{\prime}+\mathbf{q}-\mathbf{k}}c^{\mathbf{q},\mathbf{k},\mathbf{k^{\prime}}}_{eh})
	\end{multline}

	As the $c$ and $c^{\prime}$ bands are degenerate, they are henceforth referred to by ``$e$'' and the valence band by ``$h$''. In an excitonic gas, we neglect electron-electron and hole-hole correlations and set  $c^{\mathbf{q},\mathbf{k},\mathbf{k^{\prime}}}_{ee} = c^{\mathbf{q},\mathbf{k},\mathbf{k^{\prime}}}_{hh} = 0$ \cite{24}.

	\begin{figure}
		\centering
		\includegraphics[scale=0.5]{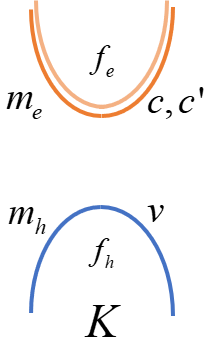}
		\caption{Reduced Model of the System}
		\label{fig5}
	\end{figure}

\end{flushleft}

%------------------------------------------------

\subsection{Conservation Laws}
Using a conservation law for the total number of carriers $N$ in the system,
\begin{equation*}
\sum_{\mathbf{k}^{\prime},\gamma^{\prime}}\braket{a^{\dagger}_{\gamma,\mathbf{k}}a^{\dagger}_{\gamma^{\prime},\mathbf{k^\prime}}a_{\gamma^{\prime},\mathbf{k^\prime}}a_{\gamma,\mathbf{k}}} = (N-1)f^{\gamma}_{\mathbf{k}}
\end{equation*}

we arrive at these relations which connect the electron/hole distributions with the electron-hole correlations:

$$
4(f^{e}_{\mathbf{k}})^2-2f^{e}_{\mathbf{k}}-c^{0,\mathbf{k},\mathbf{k^{\prime}}}_{eh} = 0,\, \qquad
(f^{h}_{\mathbf{k}})^2-f^{h}_{\mathbf{k}}-c^{0,\mathbf{k},\mathbf{k^{\prime}}}_{eh} = 0. 
$$

These equations force $f^{h}_{\mathbf{k}} = 2f^{e}_{\mathbf{k}}$.
%------------------------------------------------
\subsection{The Density Dependent Wannier Equation}

In this model, we only consider correlated e-h pairs or excitons at rest. So we make the substitution $c^{\mathbf{q},\mathbf{k},\mathbf{k^{\prime}}}_{eh} = \delta_{\mathbf{q},0}g^{eh}_{\mathbf{k},\mathbf{k^{\prime}}} = \delta_{\mathbf{q},0} \phi^{*}_{\mathbf{k}}\phi_{\mathbf{k}^{\prime}}$, where $g^{eh}_{\mathbf{k},\mathbf{k^{\prime}}}$ is a pair-correlation function and $\phi_{\mathbf{k}}$ is the Fourier transform of the excitonic wavefunction.

With the constraint of a fixed number of carriers, $N = \sum_{\mathbf{k}}f^{e}_{\mathbf{k}}$, we minimize the functional, $E_{MB} - \lambda N$, where $\lambda$ is a Lagrange multiplier. Utilising the conservation relations, we arrive at the following Wannier equation for excitons:
\begin{equation}
\alpha_{\mathbf{k}}\phi_{\mathbf{k}}+8\Delta_{\mathbf{k}}(f^{e}_{\mathbf{k}}-\frac{1}{4}) = 0
\end{equation}

where $\alpha_k = \frac{2\hbar^2\mathbf{k}^2}{\mu}-8\sum_{\mathbf{k},\mathbf{k}^{\prime}}V_{\mathbf{k}-\mathbf{k}^{\prime}}f^{e}_{\mathbf{k}^{\prime}}-\lambda$, ($\mu^{-1} = m^{-1}_e+m^{-1}_h$ being the reduced mass) and $\Delta_k = \sum_{\mathbf{k}^{\prime}}V_{\mathbf{k}-\mathbf{k}^{\prime}}\phi^{*}_{\mathbf{k}^{\prime}}$.

Starting with an initial guess for $f^{e}_{\mathbf{k}}$ and $\phi_{k}$, and defining $\Omega_k = \sqrt{\alpha_k^2+16\Delta_k^2}$ using the equations $f^{e}_{\mathbf{k}} = \frac{1}{4}(1-\frac{\alpha_k}{\Omega_k})$, and $\phi_k = \frac{2\Delta_k}{\Omega_k}$, we iteratively solve the system to obtain the excitonic wavefunctions and carrier distributions. Note that the constructed equations satisfy both the Wannier equation and the conservation relations. It takes about 20 iterations to reach $1\%$ error in the results.

\subsection{Results and Comparison}

Binding energies for excitons are obtained as a function of carrier density by subtracting the many body energy per particle from the energy for an uncorrelated equivalent system of charge carriers. All required parameters of the theory are taken from ab-initio studies\cite{RN115,RN116}.
Fig.~\ref{fig5}(a) shows the variation in binding energy for excitons in $\text{WS}_2$ monolayers over a variety of substrates. We find qualitatively similar behaviour of decreasing binding energy with increasing carrier density for $\text{WS}_2$ on a Sapphire substrate in Ref\cite{RN113} . While the model does not calculate the renormalized bandgap, the binding energies do exhibit a reduced rate of change as the Mott transition is reached. We predict a Mott transition at around $10^{14}\,\text{cm}^{-2}$ for SiO2-supported samples and  $4\times10^{13}\,\text{cm}^{-2}$ for Sapphire supported substrates, consistent with reported values\cite{RN112}.

\begin{figure}[h]
	\centering
	\includegraphics[scale=0.3]{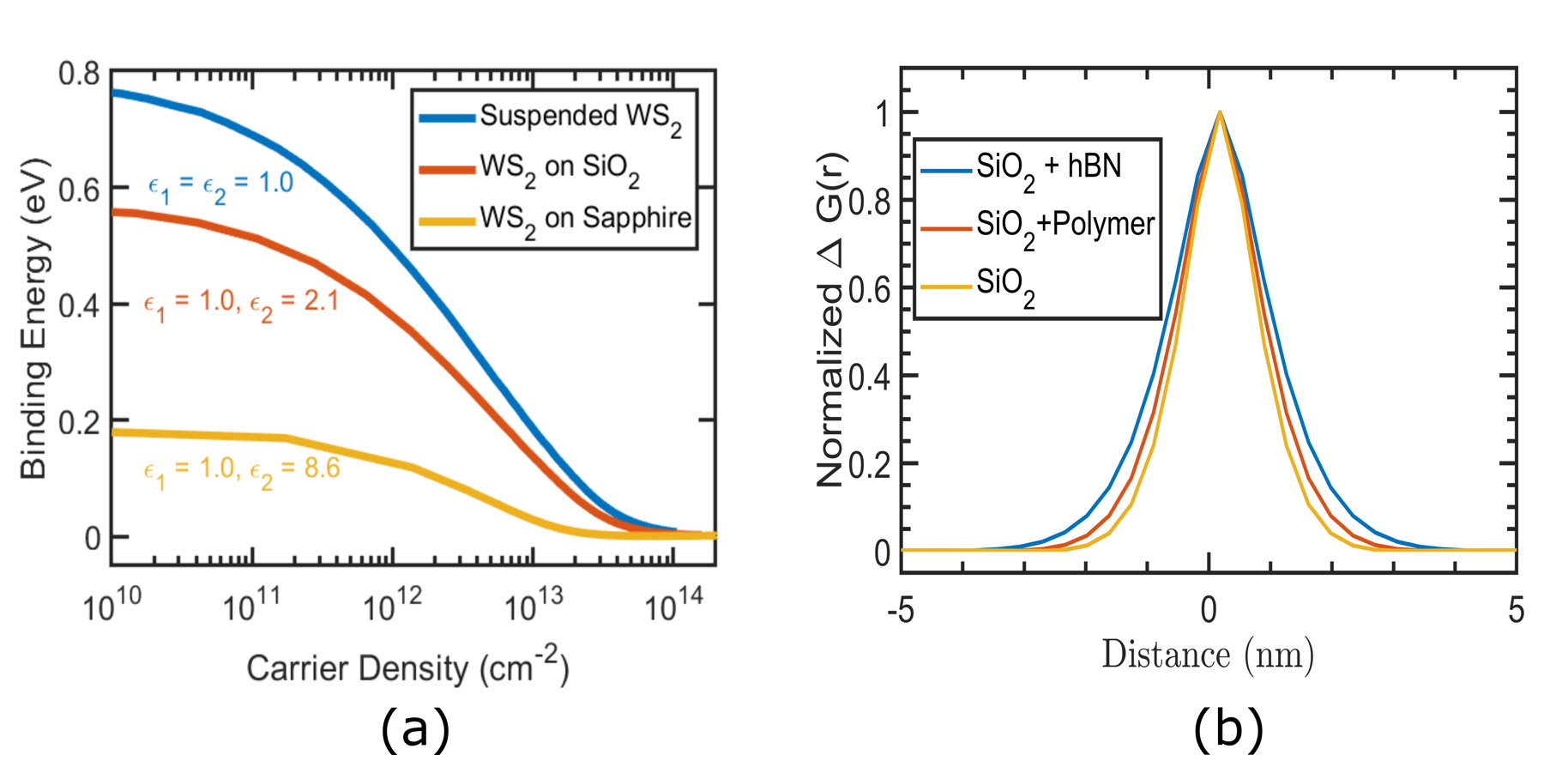}
	\caption{(a) Exciton binding energy in $\text{WS}_2$ as a function of carrier density for different substrates, (b) exciton radius for different substrates for $\text{WSe}_2$.}
	\label{fig5}
\end{figure}

The excitonic wavefunctions obtained provide an estimate of the radius of the exciton. At carrier densities of $10^{12}\,\text{cm}^{-2}$, we find good qualitative agreement with experimentally reported values of exciton radii, acquired from diamagnetic shifts in excitonic energies under high magnetic fields\cite{RN114}. In Fig.~\ref{fig5}(b) we see that the model accurately captures the spreading of the excitonic radius with an increase in dielectric screening by the environment. The plot shows the normalized pair-correlation function, $\Delta G(\textbf{r}) = |\phi(\textbf{r})|^2/|\phi(\textbf{0})|^2$, where $\phi(\textbf{r})=\frac{1}{S}\sum_{\textbf{k}}\phi_{\textbf{k}}e^{i\textbf{k}.\textbf{r}}$ and $S$ is the sample area under consideration.
We have seen that the simple model developed here captures most of the essential spin-free physics of excitons in TMDC monolayers. The absence of the need for any fitting parameters in obtaining qualitative and quantitative agreement with the results of other experiments comes at the cost of extreme limitation of the model. The model describes a uniform gas of 1s excitons without any support for treating other multi-electronic complexes, dark excitons and inter-valley scattering, which more sophisticated models based on the Wannier-Dirac Equation, or ab-initio GW, BSE-based approaches can treat. 
\bibliography{achemso-demo}